# Postoperative glioblastoma segmentation: Development of a fully automated pipeline using deep convolutional neural networks and comparison with currently available models

*Santiago Cepeda [1], Roberto Romero [2,3], Daniel García-Pérez [4], Guillermo Blasco [5], Luigi Tommaso Luppino [6], Samuel Kuttner [6,7], Ignacio Arrese [1], Ole Solheim [8,9], Live Eikenes [10], Anna Karlberg [10,11], Ángel Pérez-Núñez [12,13,14], Trinidad Escudero [15], Roberto Hornero [2,3,16], Rosario Sarabia [1]*

[1] Department of Neurosurgery, Río Hortega University Hospital, 47014 Valladolid, Spain
[2] Biomedical Engineering Group, Universidad de Valladolid, 47011 Valladolid, Spain
[3] Center for Biomedical Research in Network of Bioengineering, Biomaterials and Nanomedicine (CIBER-BBN), 47011 Valladolid, Spain
[4] Department of Neurosurgery, Albacete University Hospital, 02008, Albacete, Spain
[5] Department of Neurosurgery, La Princesa University Hospital, 28006, Madrid, Spain
[6] Department of Physics and Technology, UiT The Arctic University of Norway, 9019 Tromsø, Norway
[7] The PET Imaging Center, University Hospital of North Norway, 9019 Tromsø, Norway
[8] Department of Neurosurgery, St. Olavs University Hospital, 7030 Trondheim, Norway
[9] Department of Neuromedicine and Movement Science, Norwegian University of Science and Technology, 7034 Trondheim, Norway
[10] Department of Circulation and Medical Imaging, Faculty of Medicine and Health Sciences, Norwegian University of Science and Technology (NTNU), 7034 Trondheim, Norway
[11] Department of Radiology and Nuclear Medicine, St. Olavs Hospital, Trondheim University Hospital, 7030 Trondheim, Norway
[12] Department of Neurosurgery, 12 de Octubre University Hospital (i + 12), 28041 Madrid, Spain
[13] Department of Surgery, School of Medicine, Complutense University, 28040 Madrid, Spain
[14] Instituto de Investigación Sanitaria, 12 de Octubre University Hospital (i + 12), 28041 Madrid, Spain
[15] Department of Radiology, Río Hortega University Hospital, 47014 Valladolid, Spain
[16] Institute for Research in Mathematics (IMUVA), University of Valladolid, 47011 Valladolid, Spain

## ABSTRACT

**Background**

The pursuit of automated methods to assess the extent of resection (EOR) in glioblastomas presents a challenging endeavor, demanding precise measurement of the pretreated tumor and the residual tumor volume. Many algorithms have focused primarily on preoperative scans, making them unsuitable for postoperative studies, where the considerable variability in image patterns often hampers interpretation. Our objective was to develop a pipeline that integrates the processing of multiparametric magnetic resonance imaging (MRI) and the automatic segmentation of tumor subregions in postoperative scans using convolutional neural networks. Furthermore, we compare the performance of our pipeline against other currently available methods.

**Methods**

For the development of the segmentation model, a diverse training cohort of glioblastoma patients from three collaborative research institutions and three public databases was used. Multiparametric MRI scans were utilized as input, with ground truth labels covering contrast-enhancing tumor (ET), surgical cavity (CAV), or necrosis and edema (ED).

The models were trained using the MONAI and nnU-Net frameworks. Evaluation was performed on an external cohort from two Spanish centers and one public database, employing Dice score, Jaccard similarity coefficient, and volumetric similarity index metrics for comparison. Additionally, the model's capacity to classify the EOR was assessed, with a $< 0.1$ cm$^3$ threshold distinguishing gross total resection (GTR) from residual tumor (RT).

Finally, a comparison was made with other postoperative segmentation models, including PICTURE-nnU-Net, HD-GLIO, DeepEOR, Raidionics AGU-Net, and Emory University's model.

**Results**

A total of 184 patients and 395 MRI scans were included for model training. Among these scans, 182 were acquired within 72 hours post-surgery, 102 were follow-up scans, and 112 were preoperative scans. The external validation cohort comprised 52 scans, consisting of 36 early postoperative scans and 16 late postoperative scans.

The nnU-Net framework yielded the best model, with Dice scores of 0.761 for ET, 0.716 for CAV, and 0.734 for ED. Furthermore, the best trained model successfully categorized patients based on EOR into two groups, GTR and RT, achieving F1 scores of 91% and an accuracy of 92%.

**Conclusion**

Based on our findings, our segmentation model demonstrates a performance on par with the top-performing models currently available. Notably, our model exhibits superior precision in classifying the EOR compared to existing alternatives. By consolidating all image processing and segmentation tasks into a unified pipeline, we developed a freely accessible tool with promising clinical applicability.

**Keywords:** glioblastomas, segmentation, postoperative, neural network, deep learning

# INTRODUCTION

Glioblastoma is the most prevalent malignant brain tumor and is characterized by a dismal prognosis with a median overall survival (OS) of approximately 15 months [1]. The extent of resection (EOR) is linked to survival, as recognized in various studies [2–5].

The quantification and classification of patients based on their EOR not only holds therapeutic and prognostic significance but also serves as a critical criterion for eligibility or stratification factor in clinical trials. Magnetic resonance imaging (MRI) is the preferred imaging modality for characterizing and monitoring these tumors. Postoperative MRI, which is specifically recommended within 72 hours following surgery, plays a pivotal role in estimating the residual volume of contrast-enhancing tumor, aiding in the assessment of EOR [6–8].

Automating the segmentation of residual tumor and assessing the EOR substantial challenges for radiologists, particularly in postoperative studies where issues such as hemorrhagic debris, ischemic changes, and artifacts are common. Inter-rater agreement of manual tumor segmentations is excellent before surgery, but poor immediately after surgery and at progression. According to previous publications, the median interquartile range of EOR among raters is 8% [9]. Thus, central review of images is often necessary in multicenter clinical trials. Precise and robust segmentation of the residual tumor and surgical cavity is vital for optimal radiation treatment planning. Consequently, there is increasing interest in developing methods to automate these tasks, with neural networks emerging as a promising approach for the automatic segmentation of residual tumor and surgical cavity, as highlighted in various publications [10–20].

Our objective is to utilize two convolutional neural network (CNN) frameworks known for their robustness in medical image segmentation tasks: MONAI (https://monai.io/) and nnU-Net [21] (https://github.com/MIC-DKFZ/nnUNet). We aim to use these architectures to segment the residual enhancing tumor, the peritumoral region, and the postsurgical cavity. We hypothesize that effective training of a postoperative segmentation model requires a diverse sample encompassing preoperative, early postoperative, and follow-up studies. A multi-institutional cohort is essential to ensure variability in acquisition protocols and scanner manufacturers, as well as to accurately represent the various categories of EOR encountered in clinical practice.

We plan to use other pre-trained and publicly available state-of-the-art tumor segmentation algorithms to make inferences on a external validation cohort and compare their performance with the models trained using our dataset. In pursuit of a method suitable for longitudinal scans, we also intend to evaluate the model's applicability in preoperative scans. We delve into the strengths and limitations of each algorithm, thus charting the course for future advancements in this domain.

Our main contribution lies in the development of a publicly accessible pipeline that integrates multiparametric MRI preprocessing with an automatic segmentation method, encompassing all tumor subregions, including the postoperative cavity. Furthermore, to the best of our knowledge, there exists no published comparison of existing methods for segmenting postoperative scans in glioblastomas, a gap we aim to address through our study.

## METHODS

*Dataset description*

The training dataset consisted of a multi-institutional cohort of patients who underwent surgery with a confirmed pathological diagnosis of IDH-wildtype glioblastoma according to the latest 2021 WHO Classification of Tumors of the Central Nervous System [22]. A total of 184 patients and 395 scans constituted the training cohort, distributed as follows: 57 patients from the Río Hortega University Hospital, Valladolid, Spain; 33 patients from St. Olavs University Hospital, Trondheim, Norway; 38 patients from The LUMIERE Dataset [23]; 30 patients from Burdenko's Glioblastoma Progression Dataset [24,25]; 21 patients from the 12 de Octubre University Hospital, Madrid, Spain; and 5 patients from the Ivy Glioblastoma Atlas Project (IvyGAP) dataset [26,27]. For each included patient, the following MRI sequences were employed: T1-weighted (T1w), contrast-enhanced T1-weighted (T1ce), T2-weighted (T2w), and fluid attenuated inversion recovery (FLAIR) images. Patients with inadequate image quality or missing MRI sequences were excluded from the study.

Regarding the timing of the MRI studies, the training cohort included 181 early postoperative scans (EPSs), defined as those conducted within the initial 72 hours following surgery, in accordance with current guidelines [6–8]. Additionally, the training cohort included 112 preoperative scans and 102 follow-up scans, where tumor recurrence was diagnosed based on the modified RANO criteria [28].

The external validation cohort comprised two Spanish centers and one public dataset. To account for differences in postoperative time acquisition and EOR and for evaluation purposes, we categorized the cohort into two subgroups.

The first subgroup, from Spanish institutions, included EPS of glioblastomas treated with complete and partial resection. The study included 15 patients from La Princesa University Hospital, Madrid, Spain, and 21 patients from Albacete University Hospital, Spain.

The second subgroup comprised late postoperative scans (LPSs), encompassing patients who were scanned after surgery but before the initiation of radiation therapy, with a range of 2-5 days between scans. This subgroup consisted of 16 patients from the Quantitative Imaging Network Glioblastoma (QIN-GBM) Treatment Response dataset [25,29,30], all of whom underwent partial tumor resection.

Finally, we utilized the online validation dataset BraTS'20 (https://ipp.cbica.upenn.edu/) assess the model's performance on preoperative scans. This dataset comprises 125 patients, and detailed descriptions can be found in the associated publications. [31–33].

The distribution of time point scans and their characteristics are outlined in Table 1. The acquisition protocols for each of the sample centers are provided in the supplementary materials.

The utilization of anonymous data was authorized by the Regional Committee for Medical and Health Research Ethics (REK), Norway, and the Research Ethics Committee (CEIm) at the Río Hortega University Hospital, Valladolid, Spain, with approval number 21-PI085.

*Ground truth segmentation*

All ground truth segmentations of the training dataset and external validation datasets were conducted by two neurosurgeons (S.C. and I.A.) with over 10 years of experience in neuroimaging of brain tumors. ITK-SNAP software, version 4.0.1 (http://itksnap.org), was utilized for this task. Initially, semiautomatic segmentation was performed using the active contour tool and the clustering mode. Three labels were generated:

- Label 1 - Contrast-enhancing tumor (ET): A residual tumor was defined as enhancing tissue in T1ce but concurrently hypointense in the T1w sequence to differentiate it from residual blood, as blood appears hyperintense in both T1w and T1ce.
- Label 2 - Edema/infiltration (ED): Defined as all peritumoral T2-FLAIR signal alterations.
- Label 3 - Surgical cavity (CAV): The segmentation of the surgical cavity included hematic debris, hemostatic material, and air.

Subsequently, each label was manually corrected slice by slice. For preoperative studies, label 3 was assigned to necrosis (NEC). For follow-up studies, label 3 included both CAV and NEC if both were identifiable. The segmentations were reviewed and approved by a neuroradiologist (T.E.) with over 15 years of experience. The approximate segmentation time for each patient was 35 minutes.

*Image preprocessing*

Multiparametric MRI scans were converted to the Neuroimaging Informatics Technology Initiative (NifTI) format using dcm2niix (https://github.com/rordenlab/dcm2niix). Following this, they were coregistered to the SRI24 anatomical atlas. [34] space and resampled at $1 \times 1 \times 1$ mm isotropic voxel resolution using SimpleElastix [35]. Then, skull stripping was performed using a SynthStrip (https://github.com/freesurfer/freesurfer/tree/dev/mri_synthstrip) [36]. Finally, intensity normalization was performed using the Z score method. The final dimensions of the images were set to 240 x 240 x 155 voxels. The entire processing pipeline is openly accessible through the repository https://github.com/smcch/Postoperative-Glioblastoma-Segmentation. For datasets sourced from public repositories, the processing pipeline was tailored to meet the specific requirements of each dataset, incorporating only the essential steps, if needed, for each case. Additionally, attention was given to the variations in labels among different algorithms, ensuring their comparability with those of the ground truth. The preprocessing requirements for each model included in the comparison were properly fulfilled.

*MONAI framework training description*

We used the UNETR network architecture [37] within the MONAI framework, focusing on technical specifics to optimize performance. MRI volumes were resized to 128 x 128 x 64 voxels. The data augmentation pipeline included random flips, rotations, elastic deformations, and intensity adjustments. UNETR was configured with 4 input and 4 output channels (including background), a feature size of 32, a hidden size of 768, 12 attention heads, and a DiceFocal loss function. The model was trained for 200 epochs per fold. An ensemble evaluation of models from different folds was used to finalize segmentation predictions, utilizing a voting mechanism to improve accuracy. Postprocessing techniques or refinement of the predicted segmentations were not used. The model trained using this framework was named: the Río Hortega Glioblastoma Segmentation UNETR (RH-GlioSeg-UNETR).

*Description of the training process using nnU-Net framework*

We used the nnU-Net framework in its 3D full-resolution version, using a dataset partitioned into 5 folds for cross-validation, with each fold trained over 1000 epochs. The loss function combined Dice and cross-entropy. Data augmentation techniques such as rotations, scaling, Gaussian noise and blur, brightness and contrast adjustments, low-resolution simulations, gamma correction, and mirroring were applied to enhance the robustness of the model. This setup was designed to achieve precise segmentation results through detailed feature extraction and extensive model training. Using this framework, no postprocessing techniques were applied to the predicted segmentations. The model trained using this framework was named: the Río Hortega Glioblastoma Segmentation UNETR (RH-GlioSeg-nnU-Net).

*Extent of resection definition*

Postoperative scans were analyzed individually, without an associated preoperative study, and classified by expert observers as GTR or residual tumor (RT). GTR was defined as the absence of contrast-enhancing RT or a volume less than 0.1 $cm^3$. This strict threshold has been adopted in line with similar studies, taking into account factors such as the size of the voxel, the minimum size interpretable by the human eye, and the necessity to differentiate RT from small linear enhancements of pia matter in the walls of the surgical cavity and small blood vessels [38]. Patients whose scans failed to meet these criteria were classified as having RT.

*Evaluation metrics*

To evaluate the trained models' performance, we used overlap metrics, such as the Dice score and the Jaccard similarity coefficient (JSC), calculated using both the ground truth labels and the predicted segmentations. Additionally, the volumetric similarity index was computed, considering the significance of the predicted volume in this type of segmentation task.

Considering that postoperative studies include patients who underwent GTR, there is a unique circumstance where the absence of this label in the ground truth limits the application of overlap metrics. Therefore, following a previous study by Helland et al. [19], we divided the patients in the external validation cohort with early EPS into two subgroups. In the first group, referred to as 'Positive patients', patients had residual tumor in the ground truth, while the other group was termed 'True Positives' when, in addition to having residual tumor in the ground truth, there was also residual tumor in the predicted label.

To assess the models' ability to classify scans into GTR and RT, precision, recall, F1 score, and accuracy were employed. A threshold of 0.1 $cm^3$ was utilized for both the ground truth segmentation and EOR classification to categorize the aforementioned groups [38].

*Models used for comparison*

The main automatic segmentation models currently available were used. They were the following:

- *DeepMedic (https://github.com/deepmedic/deepmedic)*: a deep learning model designed for brain lesion segmentation from 3D MRI that utilizes a dual-pathway 3D convolutional neural network (CNN) architecture with 11 layers. This design incorporates multiscale processing through parallel pathways: one for original resolution data to capture detailed spatial information and another for downsampled data to grasp broader context. The integration of these pathways allows for effective segmentation of lesions varying in size. The model employs 3D convolutional kernels, enhancing its ability to leverage volumetric data. DeepMedic is optimized using a combination of Dice coefficient loss and cross-entropy loss, addressing class imbalances prevalent in medical imaging. A 3D fully connected conditional random field postprocessing step refines segmentation outcomes, improving boundary delineation. Training is supported by data augmentation techniques such as rotation, scaling, and mirroring, ensuring the robustness and generalizability of the model. Despite not being specifically designed for postoperative studies, as it is considered one of the state-of-the-art segmentation methods, it was included in this comparison. The output of the model categorizes the segmented volumes into three labels: edema, necrosis and enhancing tumor residue.

- *HD-GLIO (https://github.com/NeuroAI-HD/HD-GLIO)* [39,40]: A model specifically designed for the segmentation of glioma components in brain scans, distinguishing between enhancing and nonenhancing tumor regions. Trained on a dataset of longitudinal MR images, HD-GLIO leverages deep learning architectures to process spatial and temporal information, enabling it to capture the evolution of tumor characteristics over time. This model is particularly adept at segmenting nonenhancing tumor residues, a critical aspect for treatment planning and monitoring. HD-GLIO's output includes detailed volume labels for both nonenhancing and enhancing tumor residues, facilitating precise assessments of tumor progression or regression in response to therapy.

- *PICTURE nnU-Net (https://gitlab.com/picture-production/picture-nnunet-package)* [19,39,40]: an adaptation of the nnU-Net framework for medical image segmentation by incorporating pretrained models and advanced data augmentation techniques. The 'postop_beta' model integrates capabilities from two distinct models: one trained on high-grade gliomas focusing on cavities and enhancing tumor remnants and the HD-GLIO model, developed for longitudinal scans to identify nonenhancing tumor. The output of the model categorizes the segmented volumes into three labels: nonenhancing tumor residue, cavity, and enhancing tumor residue, offering a comprehensive approach to tumor segmentation.

- *DeepEOR* [13]: This model is based on the U-Net architecture. For the postoperative segmentation model, training was conducted with 72 MRI scans, and validation was performed on 45 scans drawn from the BraTS 2015 dataset [31] and the Zurich collection. The output labels for postoperative scans included enhancing the residual tumor and edema but not the surgical cavity.

- *Emory University* [10]: This model was designed for postsurgical brain tumor segmentation for radiation treatment planning and longitudinal monitoring using deep learning. By leveraging a dataset of 225 glioblastoma patients, various models, including U-Net, ResUnet, and 3D U-Net, were compared, and 3D U-Net was ultimately identified as the most effective model. The labels are the contrast-enhancing tumor, edema and surgical cavity.

- *Raidionics AGU-Net ([https://github.com/raidionics/Raidionics](https://github.com/raidionics/Raidionics))* [41,42]: Raidionics is an open-source software developed to address the need for standardized and automatic tumor segmentation and the generation of clinical reports for patients with central nervous system tumors. It offers a user-friendly graphical user interface and a robust processing backend, supporting preoperative segmentation for glioblastomas, lower-grade gliomas, meningiomas, and metastases, as well as early postoperative segmentation for glioblastomas.

For both training and evaluation of the models, a machine equipped with an Intel Core i7 processor, 64 GB of RAM, and a dedicated RTX 3090 24 GB GPU was utilized. The model based on the MONAI framework and nnU-Net was trained using Python 3.9 and PyTorch version 2.1.1 + cu121. For the Emory University and DeepEOR models, TensorFlow version 2.10.0 was employed. Raidionics AGU-Net was executed via its graphical interface on the Windows 10 operating system. PICTURE-nnU-Net, HD-GLIO, and DeepMedic were implemented in WSL Ubuntu version 20.04.4 LTS using Python 3.8, TensorFlow version 2.13.0, and PyTorch version 2.0.1.

## RESULTS

Across the entire validation dataset, the top-performing model trained with the nn-U-Net framework (RH-GlioSeg-nnU-Net) achieved Dice scores of 0.60, 0.73, and 0.72 for labels ET, ED, and CAV, respectively. An illustrative example of the predicted labels from each model included in the comparison is shown in Figure 1. A detailed comparison of the performance of the proposed algorithm with that of other available algorithms is provided in Table 2 and Figure 2A.

In the subgroup of patients from the external validation cohort with early postoperative scans (EPS), there were 23 patients who underwent GTR and 13 patients who had RT according to the ground truth labels and assessment by experts. The RH-GlioSeg-nnU-Net model achieved a Dice score of 0.61 for the residual enhancing tumor in the 'Positive' subjects group and 0.72 in the 'True Positives' subjects group. For the edema label, the Dice score was 0.75 for RH-GlioSeg-nnU-Net and 0.74 for the surgical cavity. In the 'True Positives' subjects group, the other algorithms obtained the following Dice scores for residual tumor enhancement: Picture-nnU-Net 0.59, HD-GLIO 0.62, Emory University 0.38, DeepEOR 0.19, DeepMedic 0.43, and Raidionics AGU-Net 0.50. The comparison details of the models on the subgroups of patients with EPS from the validation cohort are shown in Table 3, Figure 2B.

Using the entire validation cohort to evaluate the models' ability to accurately classify patients based on their EOR, the following accuracy values were obtained: RH-GlioSeg-nnU-Net 0.94; HD-GLIO, 0.90; Raidionics AGU-Net, 0.88; PICTURE-nnU-Net, 0.85; DeepMedic, 0.69; Emory University, 0.58; and DeepEOR, 0.56.

The results of the model evaluation on the subgroup of patients with late postoperative scans (LPS) were as follows: for label ET, the Dice scores obtained were 0.79 for RH-GlioSeg-nnU-Net, 0.67 for PICTURE-nnU-Net, 0.79 for HD-GLIO, 0.41 for Emory University, 0.73 for DeepEOR, 0.73 for DeepMedic, and 0.68 for Raidionics AGU-Net.

For the label ED, the scores were as follows: RH-GlioSeg-nnU-Net 0.71, PICTURE-nnU-Net 0.60, HD-GLIO 0.54, Emory University 0.47, DeepEOR 0.37, and DeepMedic 0.65. For the CAV label, the following Dice scores were used: RH-GlioSeg-nnU-Net, 0.67; PICTURE nnU-Net, 0.52; and Emory University, 0.36.

The details of the model comparison on the subgroups of patients with late postoperative scans from the validation cohort of patients are shown in Table 4, Figure 2C.

Figure 3 illustrates the performance of the models in segmenting enhancing residual tumor, grouped by volume, and categorized into quartiles. It shows a direct relationship between the Dice score and the volume of the ET label as determined in the ground truth.

Finally, the RH-GlioSeg-nnU-Net model attained the highest overall overlap metrics and was selected to assess its performance on preoperative MR images using the BRATS 2020 validation dataset via the online platform. The mean Dice scores obtained were 0.78, 0.88, and 0.72 for the ET, whole tumor, and tumor core labels, respectively. Details of the model evaluation on preoperative scans are provided in Table 5.

**DISCUSSION**

In this study, we compiled five datasets from collaborative research institutions and four datasets from publicly online available data sources encompassing pre- and postoperative multiparametric MRI studies. Among the postoperative scans, we distinguished between early postoperative studies and follow-up studies. Our dataset boasts diversity, stemming from multiple sources, and varying degrees of resection extension in postoperative studies. Leveraging a robust CNN architecture, we trained a model of notable reliability. Despite being primarily trained with a focus on postoperative studies, this model demonstrates dynamism and robustness, making it applicable to both preoperative and follow-up studies. This versatility is evident in the results obtained from our external validation cohorts.

Postoperative segmentation of glioblastomas presents a significant challenge, primarily due to the difficulty in accurately identifying residual enhancing tumors, especially when dealing with small volumes. The extensive variability observed in postoperative studies further complicates the standardization of methodologies. Variations in surgical techniques often result in patients exhibiting diverse EORs, despite undergoing surgery for glioblastoma in similar locations. Consequently, cases may vary from those with resections tightly confined to the enhancing component to those employing more aggressive strategies, such as supra-marginal resections or lobectomies. These differences manifest notably in terms of the size of the surgical cavity and deformation of the surrounding parenchyma. Additionally, the meticulousness of hemostasis significantly influences postsurgical outcomes, leading to clean cavities in some cases and the presence of blood debris and hemostatic material in others.

Training a model to accurately segment residual tumor, especially small volumes, poses additional challenges, particularly in reliably predicting the "absence" of residual tumor. This aspect becomes particularly crucial in classifying patients into GTR and RT categories. A model that excels at tumor segmentation may not necessarily be precise in identifying cases where no residual tumor exists, as it might tend to over segment these regions.

The ground truth poses a significant challenge in this context, with considerable inter-rater variability in EOR observed immediately after surgery, particularly when tumor volumes are small [9]. Moreover, the timing of imaging could influence results; for instance, repeat MRIs conducted only 2 days apart reveal notable variability in automatic volume measurements [43]. Additionally, the choice of software for volume segmentation may also be a critical factor to consider [44].

In our dataset, the GTR category predominated, comprising 153 scans, which accounted for 85% of all EPS. This stands in contrast to other datasets where the proportion is typically reversed. Given these circumstances, our hypothesis was that the postoperative segmentation model would derive significant benefits from learning the characteristics of the tumor both preoperatively and in follow-up studies where tumor recurrence is detected. This approach allows for achieving a balance by providing the CNN with examples of scenarios where no residual tumor is present, as well as illustrating the tumor's characteristics at different time points.

Additionally, the automation of surgical cavity segmentation has potential applications in radiotherapy treatment planning, as demonstrated by several publications. [10–12,45]. Nevertheless, there are limited models available that offer comprehensive labeling of all pertinent structures, including edema, residual tumor, and surgical cavity, specifically for postoperative studies. [10,17–20] Furthermore, not all these models are publicly accessible. The commendable endeavors undertaken by the aforementioned authors have laid the groundwork for the research presented in our work.

The rationale behind conducting a subgroup analysis of patients in the validation cohort stemmed from several factors. First, patients from Spanish centers included individuals with both GTR and RT, all with EPS, whereas subjects from the QIN-GBM dataset included only RT patients with LPS. Consequently, we anticipated potential differences in model performance based on the presence or absence of residual enhancing tumor. Notably, accurately predicting labels for GTR patients represented one of the most challenging scenarios.

By proposing this comparison, our aim was not to address criticism but rather to highlight strengths and glean insights from alternative approaches and strategies for a shared problem. Thus, we included DeepMedic, an algorithm widely regarded as part of the state-of-the-art in segmentation tasks. The model was trained on the BRATS 2015 training database, comprising 220 multimodal scans of patients with high-grade glioma (HGG) and 54 with low-grade glioma (LGG), encompassing both pre- and postoperative scans. The results from the external validation cohorts of patients with EPS demonstrated Dice scores of 0.43 and 0.61 for identifying residual tumor and edema, respectively. However, the surgical cavity label was not assessed during the evaluation, despite its similarity to the necrosis label in the originally trained model. In the cohort of patients with LPS, the scores were 0.73 and 0.65 for enhancing tumor and edema, respectively.

We also employed HD-GLIO, a model known for its robustness in postoperative dataset evaluations. HD-GLIO was trained on 3,220 MRI scans from 1,450 brain tumor patients, with only 79 being early postoperative MRI scans. Its performance is notably impressive, achieving Dice scores of 0.57 and 0.66 for residual tumor and edema, respectively, in the EPS cohort. In the LPS cohort, scores of 0.79 and 0.54 were achieved for tumor and edema segmentation, respectively.

The postoperative beta version of PICTURE-nnU-Net was also employed, notable for including the postsurgical cavity within its predicted labels, a feature that few models possess. Its performance in the EPS cohort yielded Dice scores of 0.61, 0.69, and 0.75 for residual tumor, edema, and surgical cavity, respectively. Conversely, in the LPS cohort, it achieved scores of 0.67, 0.60, and 0.52 for enhancing tumor, edema, and surgical cavity segmentation, respectively.

The subsequent two models under evaluation require only two MRI sequences for their implementation: T1ce and FLAIR. First, DeepEOR, which omits surgical cavity labels and relies on the U-Net architecture, incorporates 1053 preoperative studies and only 72 postoperative studies. Utilizing only two fundamental sequences for segmentation offers advantages in cases of missing sequences, a common occurrence in practice. However, the disparity between pre- and postoperative studies could impair its performance. In the EPS subgroup, Dice scores of 0.15 and 0.59 were achieved for residual tumor and edema, respectively. In the LPS subgroup, 0.37 and 0.37 points were assigned for enhancing tumor and edema, respectively.

Another model utilized in our study was developed by Emory University [10]. Notably, it is among the few models trained specifically to segment the postsurgical cavity, with a particular emphasis on its utility in radiotherapy planning. However, its reliance solely on T1ce and FLAIR sequences may impact its performance due to potential information gaps. In the EPS cohort, Dice scores of 0.32, 0.40, and 0.47 were achieved for the residual tumor, edema, and surgical cavity, respectively. In the LPS cohort, scores of 0.41, 0.47, and 0.36 were obtained for residual tumor, edema, and cavity, respectively.

Finally, Raidionics AGU-Net was also included in our evaluation [19,39,40]. Despite its lack of predicted labels for edema or the surgical cavity, this software offers a user-friendly graphical interface across multiple operating systems and integrates seamlessly with 3D Slicer, providing a significant advantage over other developed algorithms. Its simplicity for users makes it particularly suitable for clinical practice. In the EPS cohort, a Dice of 0.50 was achieved for residual tumors. In the LPS cohort, a score of 0.68 was obtained for residual tumors, indicating that this model ranks among the top performers for this purpose.

While striving to achieve optimal overlap metrics is understandable, it is essential to consider the significant variability among manual segmentations conducted by different observers, as demonstrated by numerous studies [9,46]. Hence, the following question arises: should we persist in striving for a flawless Dice score? Therefore, we also examined the classification of patients into EOR categories, as in clinical practice, the EOR classification has significant prognostic implications and can aid in clinical trial inclusion. The top-performing models for this task were RH-GlioSeg-nnU-Net, HD-GLIO, Raidionics AGU-Net, and PICTURE nnU-Net, which achieved accuracy values of 0.944, 0.926, 0.885, and 0.856, respectively.

It is important to emphasize that methodological comparisons among the models may not be feasible due to differences in their architectures, preprocessing and postprocessing pipelines, or the diverse datasets used for training. Therefore, our aim is not to benchmark them against each other but rather to provide a practical perspective on their performance in a clinical setting.

In terms of architectures and frameworks, we trained two models using the same dataset and employed an internal validation strategy with k-folds. However, the performance metrics are consistently higher when utilizing nnU-Net than when utilizing UNETR. Despite both being 3D fully convolutional architectures and employing similar data augmentation strategies, it appears that a more complex architecture such as UNETR does not offer significant advantages over U-Net in this specific task. Furthermore, all the models that achieved the highest scores in segmentation and EOR classification tasks were built upon the U-Net architecture.

Other noteworthy publications should be acknowledged, although they were not included in the comparison due to the unavailability of their source code for local implementation. One such example is NS-HGlio [16], which is a segmentation model with a commercial license. For training, NS-HGlio utilized studies from various centers, encompassing both pre- and postoperative data, employing the 3D U-Net architecture. In their publication, the authors reported an external validation cohort comprising 40 patients from a single site with a 1:1 ratio of pre- to postoperative scans. Their model achieved mean Dice coefficients of 0.75, 0.74, and 0.79 for enhancing tumor, edema, and whole tumor, respectively. The extent of resection (EOR) for the included patients was not specified, and the model did not incorporate segmentation of the surgical cavity.

In another paper by Nalepa et al.[17], they proposed an end-to-end pipeline for segmentation of pre- and postoperative studies. Using a multi-institutional cohort and a confidence-aware approach, nnU-Net was evaluated on a group of 40 postsurgical patients. The mean Dice coefficients for enhancing the tumor, edema, and surgical cavity were reported to be 0.63, 0.68, and 0.69, respectively. In the test group, 10 patients underwent complete resection, and only 1 patient underwent EPS; in the remaining patients, postoperative MRI were acquired up to 57 days after surgery.

The model developed by Lotan et al.[18] used pre- and postoperative studies from the 2019 BraTS dataset, employing a fusion of CNN anisotropic cascade architectures and a regularization autoencoder. With a test cohort of 40 patients, they achieved mean Dice coefficients of 0.72, 0.84, and 0.83 for enhancing tumor, core tumor, and whole tumor labels, respectively. However, the model did not incorporate segmentation of the surgical cavity, nor did it provide information on the extent of resection in the test group patients.

Finally, our model, primarily trained on early postoperative studies and follow-up data, underwent evaluation using an extensive external online validation cohort (n = 125) provided by BraTS 2020. . The mean Dice coefficients obtained were 0.78, 0.88, and 0.72 for ET, whole tumor, and tumor core labels, respectively. In comparison to previously published results [47,48], our dataset and the nnU-Net framework demonstrated remarkable performance. This highlights the versatility of our model and provides compelling evidence of its applicability to longitudinal follow-up and preoperative studies.

The limitation of our model lies in the inherent challenge of accurately segmenting postsurgical studies while encompassing all relevant regions. While manual and semiautomatic segmentation serve as standards for training and evaluation, it is essential to acknowledge the variability between observers, which introduces a bias that is difficult to eliminate.

Therefore, it is imperative to extend the evaluation of our model and others designed for this task across different settings and patient cohorts. Only through such comprehensive assessments can we ensure their reproducibility and applicability. It is important to note that these models are not intended to replace human observers but rather to enhance their efficiency and improve diagnostic precision.

In this regard, emerging initiatives like Federated Learning for Postoperative Segmentation of Treated Glioblastoma (FL-PoST) are promising solutions to address the aforementioned limitations (https://fets-ai.github.io/FL-PoST/).

## CONCLUSION

The combination of a diverse dataset from multiple institutions spanning longitudinal studies of patients with varying degrees of resection, alongside the robust framework of nnU-Net, yields exceptional outcomes in both tumor subregions and surgical cavity segmentation. Moreover, our model excels in accurately classifying the EOR across the validation cohort employed in this study. By comparing our model with existing algorithms, we can discern areas where focused efforts should be directed. This includes initiatives aimed at developing a universally accessible model that seamlessly integrates into clinical workflows. Such endeavors are pivotal for advancing the field and facilitating widespread adoption of cutting-edge segmentation and classification methodologies in clinical settings.

## ACKNOWLEDGMENTS

The authors sincerely appreciate the collaboration of all contributors to the development of the segmentation models included in this comparison. We acknowledge their technical support and willingness to share the source code of their publications. Special thanks are extended to David Bouget from the Department of Health Research, SINTEF Digital, Trondheim, Norway; Roelant S. Eijgelaar from the Neurosurgical Center Amsterdam, Amsterdam UMC, Vrije Universiteit, Netherlands; Karthik K. Ramesh from the Department of Biomedical Engineering, Emory University and Georgia Institute of Technology, Atlanta, USA; as well as Olivier Zanier and Victor E. Staartjes from the Machine Intelligence in Clinical Neuroscience (MICN) Laboratory, Department of Neurosurgery, Clinical Neuroscience Center, University Hospital Zurich, Switzerland.

## FUNDING

This work was partially funded by a grant awarded by the "Instituto Carlos III, Proyectos I-D-i, Acción Estratégica en Salud 2022", under the project titled "Prediction of tumor recurrence in glioblastomas using magnetic resonance imaging, machine learning, and transcriptomic analysis: A supratotal resection guided by artificial intelligence," reference PI22/01680.

**CONFLICTS OF INTEREST**

All authors certify that they have no affiliations with or involvement in any organization or entity with any financial interest (such as honoraria; educational grants; participation in speakers' bureaus; membership, employment, consultancies, stock ownership, or other equity interest; and expert testimony or patent-licensing arrangements), or non-financial interest (such as personal or professional relationships, affiliations, knowledge or beliefs) in the subject matter or materials discussed in this manuscript.

# REFERENCES

1. Koshy M, Villano JL, Dolecek TA, et al. Improved survival time trends for glioblastoma using the SEER 17 population-based registries. *J Neurooncol*. 2012;107(1):207-212. doi:10.1007/s11060-011-0738-7

2. Awad AW, Karsy M, Sanai N, et al. Impact of removed tumor volume and location on patient outcome in glioblastoma. *J Neurooncol*. 2017;135(1):161-171. doi:10.1007/s11060-017-2562-1

3. Gutman DA, Cooper LAD, Hwang SN, et al. MR imaging predictors of molecular profile and survival: multi-institutional study of the TCGA glioblastoma data set. *Radiology*. 2013;267(2):560-569. doi:10.1148/radiol.13120118

4. Karschnia P, Vogelbaum MA, Van Den Bent M, et al. Evidence-based recommendations on categories for extent of resection in diffuse glioma. *European Journal of Cancer*. 2021;149:23-33. doi:10.1016/j.ejca.2021.03.002

5. Karschnia P, Young JS, Dono A, et al. Prognostic validation of a new classification system for extent of resection in glioblastoma: A report of the RANO *resect* group. *Neuro-Oncology*. 2023;25(5):940-954. doi:10.1093/neuonc/noac193

6. Stupp R, Brada M, van den Bent MJ, Tonn JC, Pentheroudakis G, ESMO Guidelines Working Group. High-grade glioma: ESMO Clinical Practice Guidelines for diagnosis, treatment and follow-up. *Ann Oncol*. 2014;25 Suppl 3:iii93-101. doi:10.1093/annonc/mdu050

7. Nabors LB, Portnow J, Ammirati M, et al. NCCN Guidelines Insights: Central Nervous System Cancers, Version 1.2017. *J Natl Compr Canc Netw*. 2017;15(11):1331-1345. doi:10.6004/jnccn.2017.0166

8. Rykkje AM, Larsen VA, Skjøth-Rasmussen J, Nielsen MB, Carlsen JF, Hansen AE. Timing of Early Postoperative MRI following Primary Glioblastoma Surgery—A Retrospective Study of Contrast Enhancements in 311 Patients. *Diagnostics*. 2023;13(4):795. doi:10.3390/diagnostics13040795

9. Visser M, Müller DMJ, Van Duijn RJM, et al. Inter-rater agreement in glioma segmentations on longitudinal MRI. *NeuroImage: Clinical*. 2019;22:101727. doi:10.1016/j.nicl.2019.101727

10. Ramesh KK, Xu KM, Trivedi AG, et al. A Fully Automated Post-Surgical Brain Tumor Segmentation Model for Radiation Treatment Planning and Longitudinal Tracking. *Cancers (Basel)*. 2023;15(15):3956. doi:10.3390/cancers15153956

11. Breto AL, Cullison K, Zacharaki EI, et al. A Deep Learning Approach for Automatic Segmentation during Daily MRI-Linac Radiotherapy of Glioblastoma. *Cancers (Basel)*. 2023;15(21):5241. doi:10.3390/cancers15215241

12. Ermiş E, Jungo A, Poel R, et al. Fully automated brain resection cavity delineation for radiation target volume definition in glioblastoma patients using deep learning. *Radiat Oncol*. 2020;15(1):100. doi:10.1186/s13014-020-01553-z

13.     Zanier O, Da Mutten R, Vieli M, Regli L, Serra C, Staartjes VE. DeepEOR: automated perioperative volumetric assessment of variable grade gliomas using deep learning. *Acta Neurochir*. 2022;165(2):555-566. doi:10.1007/s00701-022-05446-w

14.     Ghaffari M, Samarasinghe G, Jameson M, et al. Automated post-operative brain tumour segmentation: A deep learning model based on transfer learning from pre-operative images. *Magnetic Resonance Imaging*. 2022;86:28-36. doi:10.1016/j.mri.2021.10.012

15.     Holtzman Gazit M, Faran R, Stepovoy K, Peles O, Shamir RR. Post-operative glioblastoma multiforme segmentation with uncertainty estimation. *Front Hum Neurosci*. 2022;16:932441. doi:10.3389/fnhum.2022.932441

16.     Abayazeed AH, Abbassy A, Müeller M, et al. NS-HGlio: A generalizable and repeatable HGG segmentation and volumetric measurement AI algorithm for the longitudinal MRI assessment to inform RANO in trials and clinics. *Neuro-Oncology Advances*. 2023;5(1):vdac184. doi:10.1093/noajnl/vdac184

17.     Nalepa J, Kotowski K, Machura B, et al. Deep learning automates bidimensional and volumetric tumor burden measurement from MRI in pre- and post-operative glioblastoma patients. *Computers in Biology and Medicine*. 2023;154:106603. doi:10.1016/j.compbiomed.2023.106603

18.     Lotan E, Zhang B, Dogra S, et al. Development and Practical Implementation of a Deep Learning–Based Pipeline for Automated Pre- and Postoperative Glioma Segmentation. *AJNR Am J Neuroradiol*. 2022;43(1):24-32. doi:10.3174/ajnr.A7363

19.     Helland RH, Ferles A, Pedersen A, et al. Segmentation of glioblastomas in early post-operative multi-modal MRI with deep neural networks. *Sci Rep*. 2023;13(1):18897. doi:10.1038/s41598-023-45456-x

20.     Bianconi A, Rossi LF, Bonada M, et al. Deep learning-based algorithm for postoperative glioblastoma MRI segmentation: a promising new tool for tumor burden assessment. *Brain Inform*. 2023;10(1):26. doi:10.1186/s40708-023-00207-6

21.     Isensee F, Jaeger PF, Kohl SAA, Petersen J, Maier-Hein KH. nnU-Net: a self-configuring method for deep learning-based biomedical image segmentation. *Nat Methods*. 2021;18(2):203-211. doi:10.1038/s41592-020-01008-z

22.     Louis DN, Perry A, Wesseling P, et al. The 2021 WHO Classification of Tumors of the Central Nervous System: a summary. *Neuro-Oncology*. 2021;23(8):1231-1251. doi:10.1093/neuonc/noab106

23.     Suter Y, Knecht U, Valenzuela W, et al. The LUMIERE dataset: Longitudinal Glioblastoma MRI with expert RANO evaluation. *Sci Data*. 2022;9(1):768. doi:10.1038/s41597-022-01881-7

24.     Zolotova SV, Golanov AV, Pronin IN, et al. Burdenko's Glioblastoma Progression Dataset (Burdenko-GBM-Progression). Published online 2023. doi:10.7937/E1QP-D183

25. Clark K, Vendt B, Smith K, et al. The Cancer Imaging Archive (TCIA): Maintaining and Operating a Public Information Repository. *J Digit Imaging*. 2013;26(6):1045-1057. doi:10.1007/s10278-013-9622-7

26. Shah N, Feng X, Lankerovich M, Puchalski RB, Keogh B. Data from Ivy GAP. Published online 2016. doi:10.7937/K9/TCIA.2016.XLWAN6NL

27. Puchalski RB, Shah N, Miller J, et al. An anatomic transcriptional atlas of human glioblastoma. *Science*. 2018;360(6389):660-663. doi:10.1126/science.aaf2666

28. Ellingson BM, Wen PY, Cloughesy TF. Modified Criteria for Radiographic Response Assessment in Glioblastoma Clinical Trials. *Neurotherapeutics*. 2017;14(2):307-320. doi:10.1007/s13311-016-0507-6

29. Mamonov AB, Kalpathy-Cramer J. Data From QIN GBM Treatment Response. Published online 2016. doi:10.7937/K9/TCIA.2016.NQF4GPN2

30. Prah MA, Stufflebeam SM, Paulson ES, et al. Repeatability of Standardized and Normalized Relative CBV in Patients with Newly Diagnosed Glioblastoma. *AJNR Am J Neuroradiol*. 2015;36(9):1654-1661. doi:10.3174/ajnr.A4374

31. Menze BH, Jakab A, Bauer S, et al. The Multimodal Brain Tumor Image Segmentation Benchmark (BRATS). *IEEE Trans Med Imaging*. 2015;34(10):1993-2024. doi:10.1109/TMI.2014.2377694

32. Bakas S, Reyes M, Jakab A, et al. Identifying the Best Machine Learning Algorithms for Brain Tumor Segmentation, Progression Assessment, and Overall Survival Prediction in the BRATS Challenge. Published online 2018. doi:10.48550/ARXIV.1811.02629

33. Bakas S, Akbari H, Sotiras A, et al. Advancing The Cancer Genome Atlas glioma MRI collections with expert segmentation labels and radiomic features. *Sci Data*. 2017;4:170117. doi:10.1038/sdata.2017.117

34. Rohlfing T, Zahr NM, Sullivan EV, Pfefferbaum A. The SRI24 multichannel atlas of normal adult human brain structure. *Hum Brain Mapp*. 2010;31(5):798-819. doi:10.1002/hbm.20906

35. Marstal K, Berendsen F, Staring M, Klein S. SimpleElastix: A User-Friendly, Multi-lingual Library for Medical Image Registration. In: *2016 IEEE Conference on Computer Vision and Pattern Recognition Workshops (CVPRW)*. IEEE; 2016:574-582. doi:10.1109/CVPRW.2016.78

36. Hoopes A, Mora JS, Dalca AV, Fischl B, Hoffmann M. SynthStrip: skull-stripping for any brain image. *NeuroImage*. 2022;260:119474. doi:10.1016/j.neuroimage.2022.119474

37. Hatamizadeh A, Tang Y, Nath V, et al. UNETR: Transformers for 3D Medical Image Segmentation. Published online October 9, 2021. Accessed December 29, 2023. http://arxiv.org/abs/2103.10504

38. Stummer W, Pichlmeier U, Meinel T, et al. Fluorescence-guided surgery with 5-aminolevulinic acid for resection of malignant glioma: a randomised controlled

multicentre phase III trial. *Lancet Oncol*. 2006;7(5):392-401. doi:10.1016/S1470-2045(06)70665-9

39. Kickingereder P, Isensee F, Tursunova I, et al. Automated quantitative tumour response assessment of MRI in neuro-oncology with artificial neural networks: a multicentre, retrospective study. *The Lancet Oncology*. 2019;20(5):728-740. doi:10.1016/S1470-2045(19)30098-1

40. Isensee F, Jäger PF, Kohl SAA, Petersen J, Maier-Hein KH. Automated Design of Deep Learning Methods for Biomedical Image Segmentation. *Nat Methods*. 2021;18(2):203-211. doi:10.1038/s41592-020-01008-z

41. Bouget D, Pedersen A, Jakola AS, et al. Preoperative Brain Tumor Imaging: Models and Software for Segmentation and Standardized Reporting. *Front Neurol*. 2022;13:932219. doi:10.3389/fneur.2022.932219

42. Bouget D, Alsinan D, Gaitan V, et al. Raidionics: an open software for pre- and postoperative central nervous system tumor segmentation and standardized reporting. *Sci Rep*. 2023;13(1):15570. doi:10.1038/s41598-023-42048-7

43. Abu Khalaf N, Desjardins A, Vredenburgh JJ, Barboriak DP. Repeatability of Automated Image Segmentation with BraTumIA in Patients with Recurrent Glioblastoma. *AJNR Am J Neuroradiol*. 2021;42(6):1080-1086. doi:10.3174/ajnr.A7071

44. Dunn WD, Aerts HJWL, Cooper LA, et al. Assessing the Effects of Software Platforms on Volumetric Segmentation of Glioblastoma. *J Neuroimaging Psychiatry Neurol*. 2016;1(2):64-72. doi:10.17756/jnpn.2016-008

45. Canalini L, Klein J, Pedrosa De Barros N, Sima DM, Miller D, Hahn HK. Comparison of different automatic solutions for resection cavity segmentation in postoperative MRI volumes including longitudinal acquisitions. In: Linte CA, Siewerdsen JH, eds. *Medical Imaging 2021: Image-Guided Procedures, Robotic Interventions, and Modeling*. SPIE; 2021:71. doi:10.1117/12.2580889

46. Tajbakhsh N, Jeyaseelan L, Li Q, Chiang JN, Wu Z, Ding X. Embracing imperfect datasets: A review of deep learning solutions for medical image segmentation. *Med Image Anal*. 2020;63:101693. doi:10.1016/j.media.2020.101693

47. Fidon L, Ourselin S, Vercauteren T. Generalized Wasserstein Dice Score, Distributionally Robust Deep Learning, and Ranger for Brain Tumor Segmentation: BraTS 2020 Challenge. In: Crimi A, Bakas S, eds. *Brainlesion: Glioma, Multiple Sclerosis, Stroke and Traumatic Brain Injuries*. Vol 12659. Lecture Notes in Computer Science. Springer International Publishing; 2021:200-214. doi:10.1007/978-3-030-72087-2_18

48. Xing Z, Yu L, Wan L, Han T, Zhu L. NestedFormer: Nested Modality-Aware Transformer for Brain Tumor Segmentation. Published online 2022. doi:10.48550/ARXIV.2208.14876

| Center/Dataset | Number of patients | Total number of scans | Preoperative | | | | Postoperative | | | | | Follow-up | | | |
|---|---|---|---|---|---|---|---|---|---|---|---|---|---|---|---|
| | | | n | Volume (cm³) | | | n | EOR (GTR/RT) | Volume (cm³) | | | n | Volume (cm³) | | |
| | | | | ET | ED | NEC | | | ET | ED | CAV | | ET | ED | NEC/CAV |
| **Training dataset** | | | | | | | | | | | | | | | |
| **All centers** | 184 | 395 | 112 | 20.39 (24.40) | 56.89 (69.96) | 9.71 (21.44) | 181 | 135/43 | 2.55 (12.31) | 29.60 (42.36) | 16.89 (21.05) | 102 | 6.72 (16.56) | 37.5 (54.21) | 8.78 (16.02) |
| Río Hortega University Hospital | 57 | 162 | 57 | 25.52 (26.66) | 62.54 (61.96) | 9.23 (15.47) | 57 | 37/17 | 0.77 (1.52) | 29.60 (35.48) | 19.60 (29.19) | 48 | 9.57 (19.33) | 54.77 (58.38) | 10.09 (20.86) |
| 12 de Octubre University Hospital | 21 | 63 | 21 | 27.85 (25.43) | 64.69 (65.35) | 10.34 (22.85) | 21 | 20/1 | 4.09 (0.00) | 34.33 (54.13) | 17.06 (22.63) | 21 | 10.18 (21.99) | 45.24 (56.05) | 10.10 (15.18) |
| St Olav's University Hospital | 33 | 87 | 29 | 16.56 (14.37) | 24.78 (47.06) | 8.67 (24.21) | 30 | 30/0 | - | 18.15 (36.02) | 10.38 (14.20) | 28 | 2.06 (4.68) | 15.03 (20.53) | 6.09 (11.40) |
| LUMIERE | 38 | 38 | - | - | - | - | 38 | 33/5 | 0.25 (0.26) | 38.17 (38.35) | 18.42 (20.67) | - | - | - | - |
| Burdenko-GBM-Progression | 30 | 30 | - | - | - | - | 30 | 10/20 | 11.62 (18.98) | 12.95 (36.37) | 15.02 (23.88) | - | - | - | - |
| Ivy-GAP | 5 | 15 | 5 | 30.40 (4.89) | 75.22 (27.56) | 21.55 (6.78) | 5 | 5/0 | - | 39.28 (41.81) | 22.80 (20.08) | 5 | 12.25 (13.39) | 55.60 (23.33) | 8.55 (5.49) |
| **External validation dataset** | | | | | | | | | | | | | | | |
| **All centers** | 52 | 52 | - | - | - | - | 52 | 23/29 | 6.08 (12.26) | 23.38 (37.11) | 20.75 (31.18) | - | - | - | - |
| **Early postoperative scan subgroup** | | | | | | | | | | | | | | | |
| Albacete University Hospital | 21 | 21 | - | - | - | - | 21 | 11/10 | 2.42 (3.50) | 23.64 (28.75) | 31.23 (38.71) | - | - | - | - |
| La Princesa University Hospital | 15 | 15 | - | - | - | - | 15 | 12/3 | 2.68 (1.61) | 34.73 (38.32) | 20.75 (31.18) | - | - | - | - |
| **Late postoperative scan subgroup** | | | | | | | | | | | | | | | |
| QIN-GBM Treatment Response | 16 | 16 | - | - | - | - | 16 | 0/16 | 13.63 (25.61) | 11.52 (38.82) | 9.27 (16.75) | - | - | - | - |

Labels: ET = residual enhancing tumor, ED = edema, NEC = necrosis, CAV = Surgical cavity. EOR = Extent of resection, GTR = Gross total resection, RT = residual tumor. *n* = number of scans. Volumes are expressed as the median and interquartile range.

| Table 2. Comparison of Model Performance Across the Entire External Validation Dataset | | | | | | | | | |
|---|---|---|---|---|---|---|---|---|---|
| Labels | Metrics | RH-GlioSeg-nnU-net | RH-GlioSeg-U-NetR | PICTURE nnU-Net | HD-GLIO | Emory University | DeepEOR | DeepMedic | Raidionics AGU-Net |
| *All subjects* | | | | | | | | | |
| ET | Dice | 0.604 (0.484, 0.725) | 0.468 (0.351, 0.584) | 0.432 (0.322, 0.543) | 0.587 (0.465, 0.709) | 0.208 (0.142, 0.275) | 0.23 (0.154, 0.306) | 0.36 (0.256, 0.464) | 0.497 (0.392, 0.602) |
| | JSC | 0.507 (0.398, 0.617) | 0.38 (0.278, 0.483) | 0.341 (0.251, 0.432) | 0.489 (0.382, 0.595) | 0.138 (0.092, 0.183) | 0.149 (0.092, 0.207) | 0.284 (0.196, 0.371) | 0.378 (0.290, 0.465) |
| | VSI | 0.676 (0.551, 0.801) | 0.562 (0.430, 0.694) | 0.523 (0.392, 0.655) | 0.728 (0.600, 0.856) | 0.32 (0.222, 0.417) | 0.329 (0.231, 0.426) | 0.413 (0.295, 0.530) | 0.59 (0.469, 0.712) |
| ED | Dice | 0.734 (0.679, 0.789) | 0.705 (0.647, 0.764) | 0.661 (0.597, 0.726) | 0.622 (0.551, 0.694) | 0.417 (0.357, 0.477) | 0.518 (0.445, 0.591) | 0.623 (0.558, 0.687) | |
| | JSC | 0.613 (0.551, 0.674) | 0.578 (0.519, 0.637) | 0.532 (0.468, 0.595) | 0.496 (0.428, 0.565) | 0.286 (0.238, 0.335) | 0.388 (0.325, 0.451) | 0.489 (0.425, 0.552) | |
| | VSI | 0.831 (0.790, 0.872) | 0.814 (0.768, 0.861) | 0.806 (0.755, 0.857) | 0.747 (0.680, 0.814) | 0.594 (0.523, 0.665) | 0.66 (0.594, 0.725) | 0.7 (0.637, 0.764) | |
| CAV | Dice | 0.716 (0.647, 0.785) | 0.674 (0.606, 0.742) | 0.683 (0.618, 0.748) | | 0.434 (0.378, 0.490) | | | |
| | JSC | 0.603 (0.534, 0.672) | 0.55 (0.484, 0.617) | 0.558 (0.493, 0.622) | | 0.298 (0.251, 0.344) | | | |
| | VSI | 0.786 (0.722, 0.850) | 0.767 (0.705, 0.829) | 0.791 (0.726, 0.856) | | 0.61 (0.536, 0.683) | | | |
| *Positive Subjects* | | | | | | | | | |
| ET | Dice | 0.708 (0.613, 0.804) | 0.661 (0.566, 0.756) | 0.641 (0.552, 0.730) | 0.689 (0.587, 0.791) | 0.374 (0.298, 0.449) | 0.27 (0.189, 0.351) | 0.596 (0.495, 0.697) | 0.566 (0.473, 0.659) |
| | JSC | 0.595 (0.500, 0.690) | 0.538 (0.443, 0.633) | 0.506 (0.427, 0.585) | 0.573 (0.480, 0.666) | 0.247 (0.191, 0.302) | 0.175 (0.113, 0.238) | 0.469 (0.375, 0.564) | 0.43 (0.348, 0.511) |
| | VSI | 0.793 (0.704, 0.882) | 0.795 (0.704, 0.886) | 0.776 (0.674, 0.878) | 0.853 (0.770, 0.936) | 0.573 (0.471, 0.675) | 0.385 (0.285, 0.485) | 0.683 (0.572, 0.794) | 0.672 (0.565, 0.778) |
| *True Positives Subjects* | | | | | | | | | |
| ET | Dice | 0.761 (0.696, 0.826) | 0.661 (0.566, 0.756) | 0.634 (0.543, 0.726) | 0.713 (0.621, 0.805) | 0.401 (0.331, 0.471) | 0.34 (0.255, 0.426) | 0.639 (0.522, 0.755) | 0.608 (0.531, 0.685) |
| | JSC | 0.639 (0.562, 0.716) | 0.538 (0.443, 0.633) | 0.499 (0.418, 0.580) | 0.593 (0.507, 0.680) | 0.265 (0.212, 0.318) | 0.223 (0.151, 0.294) | 0.508 (0.394, 0.622) | 0.461 (0.388, 0.534) |
| | VSI | 0.837 (0.771, 0.902) | 0.795 (0.704, 0.886) | 0.768 (0.664, 0.873) | 0.859 (0.774, 0.944) | 0.598 (0.496, 0.701) | 0.452 (0.342, 0.562) | 0.704 (0.576, 0.831) | 0.721 (0.636, 0.807) |
| *Gross Total resection versus Residual Tumor classification* | | | | | | | | | |
| | Precision | 0.944 | 0.854 | 0.856 | 0.926 | 0.784 | 0.279 | 0.822 | 0.885 |
| | Recall | 0.939 | 0.739 | 0.835 | 0.891 | 0.522 | 0.5 | 0.652 | 0.875 |
| | F1 Score | 0.941 | 0.738 | 0.84 | 0.899 | 0.404 | 0.358 | 0.625 | 0.879 |
| | Accuracy | 0.942 | 0.769 | 0.846 | 0.904 | 0.577 | 0.558 | 0.692 | 0.882 |

Labels: ET = residual enhancing tumor, ED = edema, CAV = Surgical cavity. Metrics: JSC = Jaccard similarity coefficient, VSI = volumetric similarity index. Segmentation metrics values are expressed as the mean and 95% confidence interval. Classification metrics are presented as average values.

| Labels | Metrics | RH-GlioSeg-nnU-net | RH-GlioSeg-U-NetR | PICTURE nnU-Net | HD-GLIO | Emory University | DeepEOR | DeepMedic | Raidionics AGU-Net |
|---|---|---|---|---|---|---|---|---|---|
| Table 3. Comparison of Model Performance on the Subgroup of Early Postoperative Scans in the External Validation Dataset ||||||||||
| *All subjects* ||||||||||
| ET | Dice | 0.441 (0.245, 0.636) | 0.248 (0.119, 0.377) | 0.293 (0.155, 0.432) | 0.41 (0.214, 0.607) | 0.11 (0.045, 0.189) | 0.054 (0.011, 0.096) | 0.168 (0.071, 0.265) | 0.322 (0.169, 0.474) |
|  | JSC | 0.365 (0.191, 0.538) | 0.191 (0.086, 0.297) | 0.226 (0.117, 0.336) | 0.336 (0.171, 0.501) | 0.078 (0.028, 0.128) | 0.033 (0.005, 0.060) | 0.121 (0.048, 0.194) | 0.228 (0.115, 0.341) |
|  | VSI | 0.495 (0.299, 0.692) | 0.311 (0.155, 0.467) | 0.364 (0.196, 0.533) | 0.522 (0.337, 0.766) | 0.188 (0.080, 0.295) | 0.131 (0.043, 0.218) | 0.199 (0.084, 0.314) | 0.413 (0.212, 0.613) |
| ED | Dice | 0.746 (0.681, 0.811) | 0.703 (0.630, 0.776) | 0.687 (0.612, 0.763) | 0.66 (0.577, 0.742) | 0.395 (0.328, 0.461) | 0.585 (0.512, 0.658) | 0.613 (0.536, 0.689) |  |
|  | JSC | 0.626 (0.552, 0.700) | 0.575 (0.504, 0.647) | 0.559 (0.485, 0.633) | 0.533 (0.453, 0.614) | 0.264 (0.213, 0.315) | 0.442 (0.376, 0.509) | 0.475 (0.402, 0.548) |  |
|  | VSI | 0.831 (0.788, 0.877) | 0.801 (0.744, 0.858) | 0.822 (0.771, 0.874) | 0.763 (0.680, 0.847) | 0.544 (0.459, 0.629) | 0.704 (0.640, 0.768) | 0.672 (0.595, 0.749) |  |
| CAV | Dice | 0.737 (0.655, 0.819) | 0.67 (0.581, 0.760) | 0.75 (0.695, 0.806) |  | 0.469 (0.397, 0.540) |  |  |  |
|  | JSC | 0.627 (0.548, 0.706) | 0.552 (0.467, 0.637) | 0.623 (0.563, 0.684) |  | 0.33 (0.269, 0.390) |  |  |  |
|  | VSI | 0.806 (0.732, 0.880) | 0.753 (0.667, 0.840) | 0.838 (0.776, 0.899) |  | 0.668 (0.578, 0.758) |  |  |  |
| *Positive Subjects* ||||||||||
| ET | Dice | 0.61 (0.413, 0.807) | 0.534 (0.359, 0.709) | 0.609 (0.467, 0.752) | 0.568 (0.355, 0.781) | 0.323 (0.175, 0.472) | 0.149 (0.044, 0.254) | 0.427 (0.257, 0.596) | 0.421 (0.258, 0.583) |
|  | JSC | 0.505 (0.317, 0.692) | 0.412 (0.254, 0.569) | 0.47 (0.347, 0.593) | 0.465 (0.282, 0.648) | 0.217 (0.111, 0.322) | 0.09 (0.020, 0.161) | 0.308 (0.172, 0.443) | 0.298 (0.175, 0.421) |
|  | VSI | 0.686 (0.515, 0.858) | 0.67 (0.477, 0.863) | 0.756 (0.594, 0.918) | 0.764 (0.584, 0.944) | 0.52 (0.323, 0.717) | 0.362 (0.165, 0.558) | 0.506 (0.304, 0.708) | 0.54 (0.323, 0.756) |
| *True Positive Subjects* ||||||||||
| ET | Dice | 0.721 (0.587, 0.855) | 0.534 (0.359, 0.709) | 0.591 (0.441, 0.742) | 0.616 (0.411, 0.820) | 0.382 (0.235, 0.530) | 0.193 (0.068, 0.319) | 0.427 (0.257, 0.596) | 0.497 (0.354, 0.640) |
|  | JSC | 0.597 (0.439, 0.754) | 0.412 (0.254, 0.569) | 0.451 (0.323, 0.578) | 0.504 (0.326, 0.682) | 0.256 (0.148, 0.364) | 0.118 (0.031, 0.204) | 0.308 (0.172, 0.443) | 0.352 (0.239, 0.466) |
|  | VSI | 0.774 (0.651, 0.896) | 0.67 (0.477, 0.863) | 0.737 (0.565, 0.909) | 0.77 (0.573, 0.967) | 0.572 (0.355, 0.79) | 0.434 (0.199, 0.669) | 0.506 (0.304, 0.708) | 0.638 (0.441, 0.834) |
| *Gross Total resection versus Residual Tumor classification* ||||||||||
|  | Precision | 0.906 | 0.732 | 0.771 | 0.861 | 0.686 | 0.181 | 0.724 | 0.817 |
|  | Recall | 0.918 | 0.674 | 0.793 | 0.891 | 0.522 | 0.5 | 0.652 | 0.832 |
|  | F1 Score | 0.911 | 0.575 | 0.771 | 0.858 | 0.312 | 0.265 | 0.543 | 0.821 |
|  | Accuracy | 0.917 | 0.583 | 0.778 | 0.861 | 0.389 | 0.361 | 0.556 | 0.829 |

Labels: ET = residual enhancing tumor, ED = edema, CAV = Surgical cavity. Metrics: JSC = Jaccard similarity coefficient, VSI = volumetric similarity index. Segmentation metrics values are expressed as the mean and 95% confidence interval. Classification metrics are presented as average values.

| Table 4. Comparison of Model Performance on the Subgroup of Late Postoperative Scans in the External Validation Dataset | | | | | | | | | |
|---|---|---|---|---|---|---|---|---|---|
| Labels | Metrics | RH-GlioSeg-nnU-net | RH-GlioSeg-U-NetR | PICTURE nnU-Net | HD-GLIO | Emory University | DeepEOR | DeepMedic | Raidionics AGU-Net |
| ET | Dice | 0.788 (0.717, 0.860) | 0.76 (0.682, 0.838) | 0.667 (0.540, 0.793) | 0.787 (0.727, 0.846) | 0.414 (0.336, 0.493) | 0.368 (0.265, 0.471) | 0.733 (0.650, 0.817) | 0.684 (0.607, 0.761) |
| | JSC | 0.668 (0.580, 0.756) | 0.633 (0.535, 0.732) | 0.535 (0.421, 0.649) | 0.66 (0.584, 0.737) | 0.271 (0.209, 0.333) | 0.244 (0.154, 0.334) | 0.6 (0.501, 0.700) | 0.536 (0.451, 0.621) |
| | VSI | 0.88 (0.803, 0.957) | 0.86 (0.789, 0.930) | 0.792 (0.645, 0.938) | 0.926 (0.887, 0.964) | 0.616 (0.504, 0.729) | 0.404 (0.293, 0.516) | 0.827 (0.753, 0.901) | 0.779 (0.713, 0.845) |
| ED | Dice | 0.707 (0.591, 0.822) | 0.697 (0.587, 0.806) | 0.602 (0.471, 0.734) | 0.539 (0.393, 0.685) | 0.467 (0.333, 0.602) | 0.366 (0.206, 0.525) | 0.645 (0.512, 0.778) | |
| | JSC | 0.582 (0.460, 0.705) | 0.567 (0.450, 0.684) | 0.47 (0.342, 0.598) | 0.413 (0.278, 0.547) | 0.337 (0.224, 0.451) | 0.265 (0.135, 0.395) | 0.52 (0.381, 0.659) | |
| | VSI | 0.831 (0.735, 0.927) | 0.828 (0.742, 0.913) | 0.769 (0.642, 0.895) | 0.709 (0.585, 0.833) | 0.706 (0.580, 0.833) | 0.56 (0.400, 0.721) | 0.764 (0.645, 0.883) | |
| CAV | Dice | 0.666 (0.525, 0.806) | 0.679 (0.550, 0.809) | 0.521 (0.362, 0.681) | | 0.355 (0.276, 0.435) | | | |
| | JSC | 0.547 (0.396, 0.698) | 0.556 (0.425, 0.686) | 0.4 (0.253, 0.548) | | 0.225 (0.168, 0.282) | | | |
| | VSI | 0.739 (0.602, 0.877) | 0.765 (0.628, 0.903) | 0.678 (0.512, 0.843) | | 0.479 (0.367, 0.591) | | | |

Labels: ET = residual enhancing tumor, ED = edema, CAV = Surgical cavity. Metrics: JSC = Jaccard similarity coefficient, VSI = volumetric similarity index. Segmentation metrics values are expressed as the mean and 95% confidence interval. Classification metrics are presented as average values.

| Table 5. RH-GlioSeg-nnU-Net model performance in BraTS 2020 validation cohort. | | | | | | | | | | | | |
|---|---|---|---|---|---|---|---|---|---|---|---|---|
| | Dice_ET | Dice_WT | Dice_TC | Sensitivity_ET | Sensitivity_WT | Sensitivity_TC | Specificity_ET | Specificity_WT | Specificity_TC | Hausdorff95_ET | Hausdorff95_WT | Hausdorff95_TC |
| Mean | 0.776 | 0.884 | 0.723 | 0.781 | 0.849 | 0.677 | 1.000 | 0.999 | 1.000 | 26.806 | 5.501 | 34.412 |
| SD | 0.280 | 0.095 | 0.321 | 0.297 | 0.137 | 0.340 | 0.000 | 0.001 | 0.000 | 91.096 | 6.363 | 95.544 |
| Median | 0.880 | 0.909 | 0.895 | 0.893 | 0.904 | 0.848 | 1.000 | 1.000 | 1.000 | 2.000 | 3.606 | 3.162 |
| 25 quantile | 0.809 | 0.863 | 0.611 | 0.792 | 0.790 | 0.452 | 1.000 | 0.999 | 1.000 | 1.000 | 2.236 | 1.732 |
| 75 quantile | 0.913 | 0.941 | 0.937 | 0.949 | 0.943 | 0.937 | 1.000 | 1.000 | 1.000 | 3.000 | 5.745 | 12.247 |

Labels: ET = enhancing tumor, WT = Whole tumor, TC = Tumor core.  SD: Standard deviation

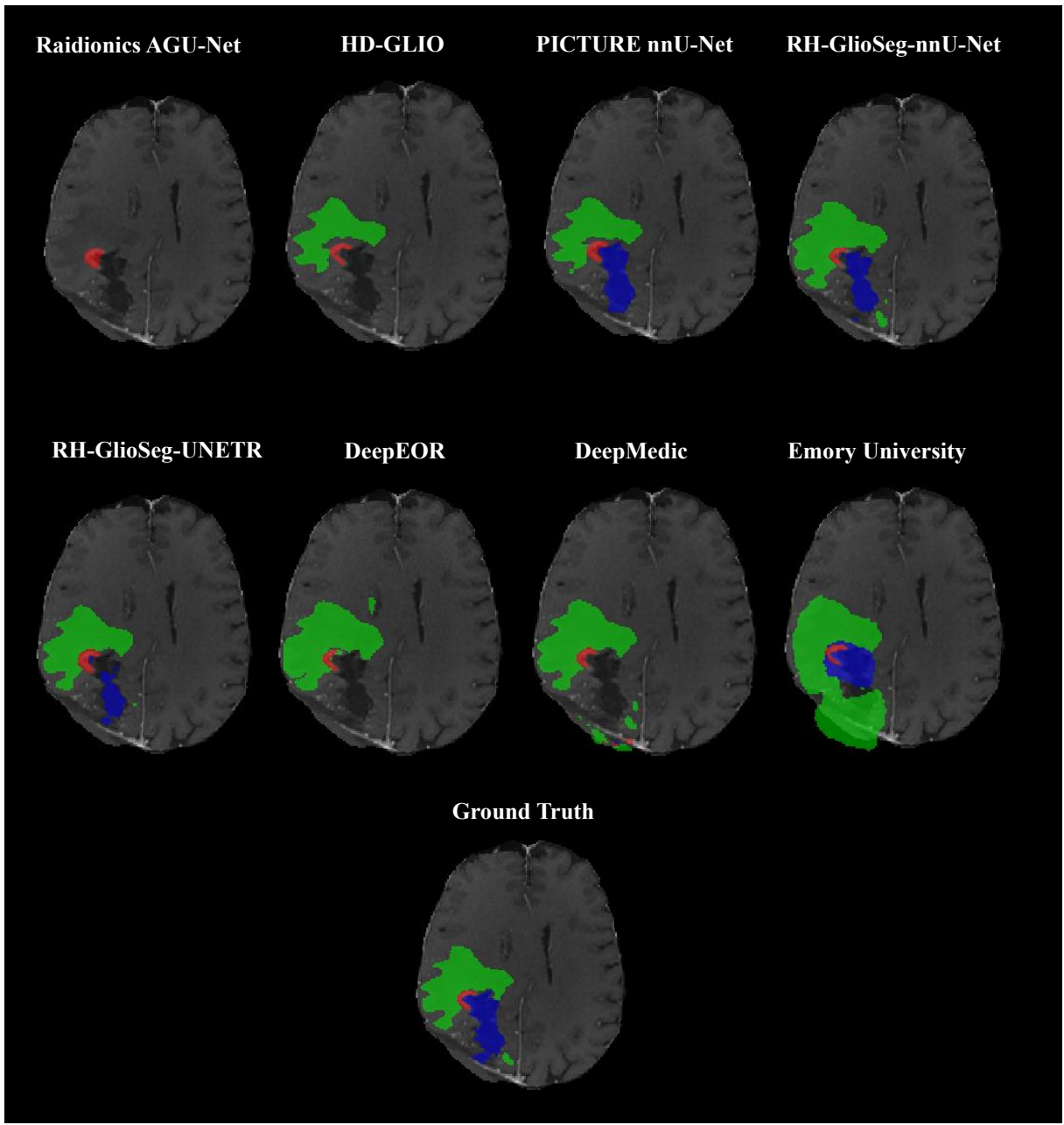

**Figure 1.** A descriptive example of the segmentations predicted by the various models included in the comparison. The segmentations include the labels: residual enhancing tumor (red), edema (green), and surgical cavity (blue).

**A**

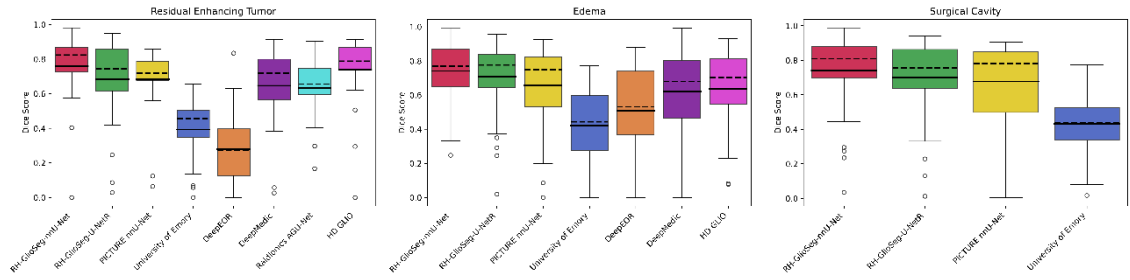

**B**

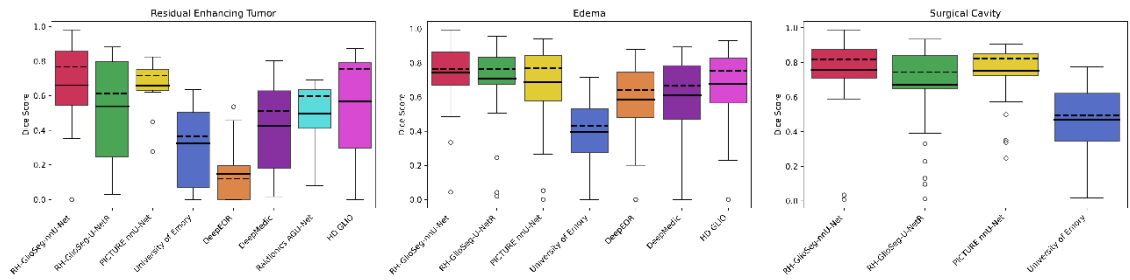

**C**

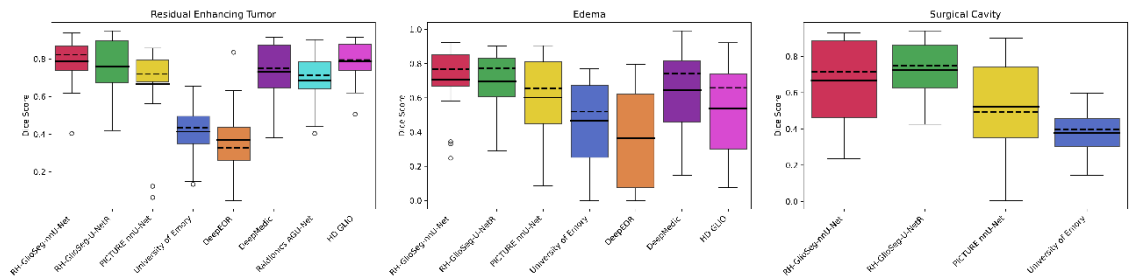

**Figure 2.** Box plot of the median (dotted line) and mean (solid line) Dice scores obtained by the evaluated models across the entire external validation patient cohort (A), early postoperative scans (EPSs) subgroup (B), and late postoperative scans (LPSs) subgroup (C).

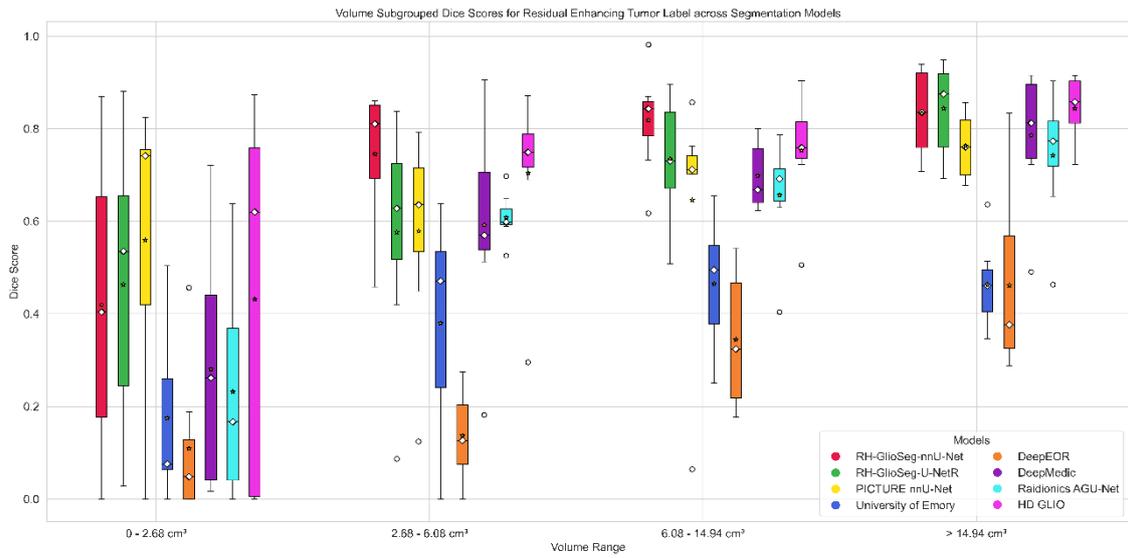

**Figure 3.** Boxplots represent the distribution of Dice scores for the residual enhancing tumor label across segmentation models, grouped by quartile-based volume divisions. The white star indicates the mean, while the white diamond represents the median.